\documentclass[aps,apl,twocolumn,superscriptaddress,showpacs,floatfix]{revtex4}
\usepackage{graphicx,epsfig,amsmath}

\newcommand{\etal}{\textit{et al}}
\newcommand{\subfig}[2]{Fig.~\ref{fig:#1}(#2)} 

\newcommand{\Vone}{\mbox{$\text{V}_{\text{1}}$}}
\newcommand{\Vtwo}{\mbox{$\text{V}_{\text{2}}$}}
\newcommand{\Tplus}{\mbox{$\text{T}_{\text{+}}$}}
\newcommand{\Tone}{\mbox{$\text{T}_{\text{1}}$}}
\newcommand{\Tm}{\mbox{$\text{T}_{\text{m}}$}}

\newcommand{\nL}{\mbox{$n_{L}$}}
\newcommand{\nC}{\mbox{$n_{C}$}}

\begin{document}

\title{Enhanced charge detection of spin qubit readout via an intermediate state}


\author{S.~A.~Studenikin}
	\affiliation{National Research Council Canada, Ottawa, ON Canada K1A 0R6}
\author{J. Thorgrimson}
	\affiliation{National Research Council Canada, Ottawa, ON Canada K1A 0R6}
	\affiliation{Department of Physics, McGill University, 3600 rue University, Montréal, QC Canada H3A 2T8}
\author{G.~C.~Aers}
	\affiliation{National Research Council Canada, Ottawa, ON Canada K1A 0R6}
\author{A.~Kam}
	\affiliation{National Research Council Canada, Ottawa, ON Canada K1A 0R6}
\author{P.~Zawadzki}
	\affiliation{National Research Council Canada, Ottawa, ON Canada K1A 0R6}
\author{Z.~R.~Wasilewski}
	\affiliation{National Research Council Canada, Ottawa, ON Canada K1A 0R6}
\author{A. Bogan}
	\affiliation{National Research Council Canada, Ottawa, ON Canada K1A 0R6}
	\affiliation{Department of Physics and Astronomy, University of Waterloo, 200 University Avenue West, Waterloo, 
	ON Canada N2L 3G1}
\author{A.~S.~Sachrajda}
  \email{Andrew.Sachrajda@nrc.ca}
	\affiliation{National Research Council Canada, Ottawa, ON Canada K1A 0R6}


\begin{abstract}

We employ an intermediate excited charge state of a lateral quantum dot device to increase the charge detection contrast during the qubit state
readout procedure, allowing us to increase the visibility of coherent qubit oscillations. This approach amplifies the coherent oscillation
magnitude but has no effect on the detector noise resulting in an increase in the signal to noise ratio. In this letter we apply this scheme to
demonstrate a significant enhancement of the fringe contrast of coherent Landau-Zener-St$\ddot u$ckleberg oscillations between singlet S and triplet
$\Tplus$ two-spin states.

\end{abstract}


\maketitle



	The crucial step of spin qubit readout via spin to charge conversion is usually achieved 
through charge detection technology.\cite{Field1993} The maximum sensitivity is related to the difference (contrast) in 
the charge detection signal of the two charge state configurations. In this letter we demonstrate how 
this contrast can be amplified by a significant factor if a ``state discriminating'' relaxation process is 
introduced during the readout procedure. Since this modification only changes the charge detector contrast 
between the relevant qubit states, the signal to noise ratio is likewise enhanced.

	Spin qubits based on single, double or triple quantum dot circuits have been successfully demonstrated.\cite{Petta2005,Petta2010,Gaudreau2011}
Following a coherent manipulation experiment the measurement is usually completed by reading out the final 
quantum state using a charge detector which is able to distinguish between the quantum spin states.

\begin{figure}
\includegraphics[width=1.0\columnwidth]{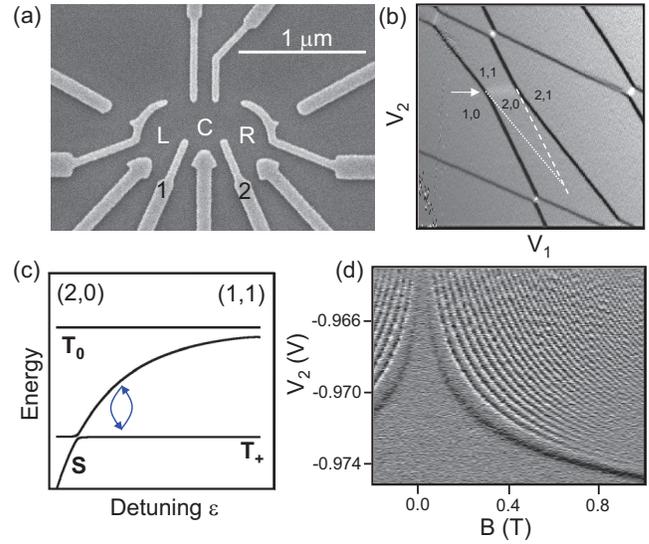} \vskip 0.0cm
\caption{(a) SEM of triple dot structure. (b) Stability diagram as a function of voltages on gates 1 and 2. The charge
transfer line between charge states (2,0) and (1,1) is indicated by the arrow. (c) Double dot energy level diagram for
dots L and C with dot R acting as a spectator. (d) Typical spin funnel and coherent LZS oscillations observed 
with the enhanced charge detection technique as a function of initial gate detuning and magnetic field B.}
 \label{fig:1}
\end{figure}

	We will concentrate on the qubit based on the singlet (S) and triplet ($\Tplus$) two-spin  states recently demonstrated by
Petta \etal.\cite{Petta2010} The corresponding oscillations are called Landau-Zener-St$\ddot u$ckleberg (LZS) oscillations and are observed
as fringes in the stability diagram or when plotted as a function of pulse duration, magnetic field or initial gate detuning. 
We stress that a similar readout advantage can be implemented for other spin qubit species. The
experiments were made using a triple quantum dot device (see \subfig{1}{a}).\cite{Gaudreau2011,Gaudreau2009,Granger2010} During our measurements one
quantum dot (R) was not coupled via exchange or tunnelling to the other two quantum dots making this effectively a two quantum dot
experiment.\cite{Studenikin2012} As a result we will use double quantum dot notation throughout. A schematic energy level diagram vs detuning is
shown for a double dot in a magnetic field in \subfig{1}{c}. The two qubit states S and $\Tplus$ which differ in spin are coupled  via the hyperfine
interaction which creates an anticrossing in the energy level diagram. 

 A generic S/$\Tplus$ spin qubit operation for this system would proceed as
follows. The quantum state  preparation is first achieved by applying a pulse (in reality a combined
pulse on gates 1 and 2) from the S(2,0) regime through the anticrossing to the (1,1) charge  regime (where ($\nL$, $\nC)$ indicate the
number of electrons, $\nL$ ($\nC$) in quantum dots L (C)). A suitable pulse rise time permits a superposition of the S and $\Tplus$ states to
be  generated via Landau-Zener tunnelling on passage through the anticrossing (in an ideal situation this would move the  state vector to
the equator of the S/$\Tplus$ Bloch sphere).\cite{Petta2010} Having created the superposition any required manipulation is performed in the
(1,1) ground state regime. In this regime and away from the anticrossing, rotations around the Z axis of the Bloch sphere (phase
accumulation during the time spent by the pulse beyond the S/$\Tplus$ anticrossing) occur with a rate dependent on the energy level
spacing (illustrated by arrows in \subfig{1}{c}). As can be seen in \subfig{1}{c} this depends on the distance from the anticrossing. The final ``readout'' process takes advantage of the
Pauli Blockade phenomenon. The system is moved back to the (2,0) ground state regime where the $\Tplus$(1,1) state can only relax to S(2,0) 
with a relaxation time $\Tone$ ($\sim$100 $\mu$s in these experiments) and coherent behaviour can be measured if $\Tone > \Tm$, the measurement time 
($\sim$10 $\mu$s). The quantum dot circuit state is typically probed by measuring the current through a nearby quantum point contact (QPC) charge
detector\cite{Field1993} or, occasionally, a neighbouring quantum dot.\cite{Barthel2009} The QPC is tuned to be highly sensitivity to local
electrostatic conditions (usually at a QPC conductance  $< G_0 = 2e^2 /h$). The (1,1) and (2,0) charge configurations possess 
different electron distributions and, therefore, have different electrostatic coupling to the charge detector. These  can be characterized
by I(2,0) and I(1,1) where I is the current through the charge detector for a fixed applied bias.  Thus the probability of the occupation
of the (2,0) state P(S)  oscillates according to the phase accumulation during the pulse and can be measured with a maximum sensitivity
contrast  equal to I(2,0) - I(1,1). To extract quantitative values for the probability is a complex procedure 
taking into account $\Tone$ processes during measurement and initialization steps. This is beyond the scope of this letter.
 
\begin{figure}
\includegraphics[width=1.0\columnwidth]{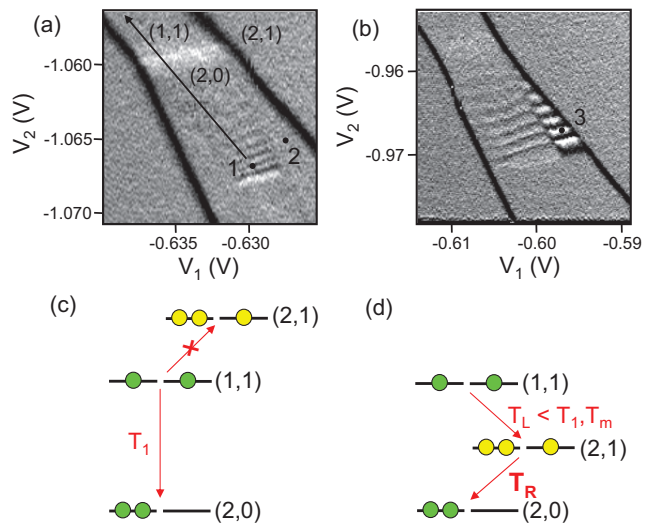} \vskip 0.0cm
\caption{(a) Expanded stability diagram showing LZS oscillations in the the charge detector current taken under conditions in which
 amplification is suppressed. $T_{R}\sim$10ns, $\Tm$=$4~\mu$s, B=100mT, pulse amplitude ($\delta$$\Vone$,$\delta$$\Vtwo$) = (-8.5,10.0)mV, rise time =
 4~ns, duration = 10~ns. Arrow indicates typical voltage pulse. (b) Same as (a) but taken under enhancement conditions. $T_{R}\sim450\mu$s,
 $\Tm$=$20~\mu$s,  B=200mT, pulse amplitude ($\delta$$\Vone$,$\delta$$\Vtwo$) = (-16.2,20.0)mV, rise time = 3~ns, duration = 7~ns.
(c) Schematic energy diagram for unamplified measurement inside triangular part of (2,0) charge region (see \subfig{1}{b}) and
corresponding to point 1 in (a). (d) Schematic energy diagram for points 2 in (a), 3 in (b) to the right of the triangular region.
Enhancement occurs if $T_{L} < \Tone , T_{R} \ge \Tm $. 
 }
 \label{fig:2}
\end{figure}

\begin{figure}
\includegraphics[width=1.0\columnwidth]{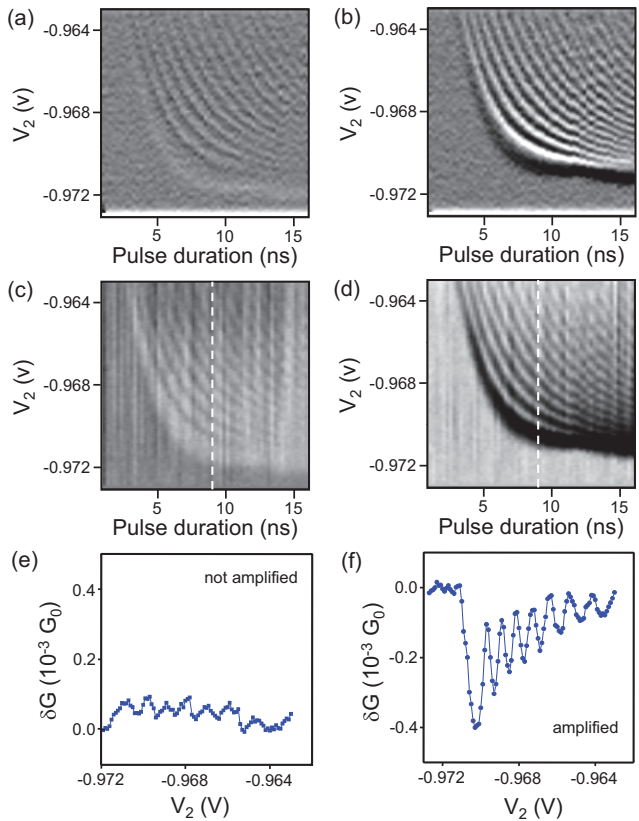} \vskip 0.0cm
\caption{(a)[(b)] Derivative of the QPC conductance with respect to initial detuning component along $\Vtwo$ as a function of
pulse duration and initial detuning for a pulse starting in the unamplified[amplified] region of \subfig{2}{b}. (c,d) Raw QPC 
conductances corresponding to (a,b). (e,f) Fixed pulse duration (9~ns) line scans taken from (c,d).
 }
 \label{fig:3}
\end{figure}

	Consider now the role of excited states in the readout measurement. The $\Tplus$(1,1) configuration is an excited 
state configuration in the (2,0) region (see \subfig{1}{b}). The dotted and dashed lines in \subfig{1}{b} mark the boundary 
along which two excited states cross. To the left (right) of the dotted (dashed) line the (1,0) ((2,1)) excited state  has a
lower energy than the (1,1) excited state. In both cases it is usually observed that the relaxation time from (1,1) to (2,0) is  very rapid in
these regions mediated first by relaxation to excited states and then by adding (removing) an electron  to (from) the spin
unpolarized leads i.e. (1,1)$\rightarrow$(1,0)$\rightarrow$(2,0) or (1,1)$\rightarrow$(2,1)$\rightarrow$(2,0). The latter
process is illustrated in \subfig{2}{d}). These processes usually involve rapid communication with the leads and this
typically restricts the useful readout region to the triangle  marked in \subfig{1}{b} where $\Tone$ can be
made larger than the measurement time required for a reasonable  measurement visibility.\cite{Johnson2005} This is
illustrated in \subfig{2}{c} for a typical pulse  measured within the triangle (point 1 in \subfig{2}{a}). Typical LZS
fringes obtained using this conventional scheme are shown in \subfig{3}{a,c,e}.

	Here we suggest and demonstrate an alternative approach taking advantage of the properties of these metastable states. This scheme leads to
dramatically enhanced LZS fringe visibility, an example of which is shown in \subfig{1}{d}. Typically about 50 coherent oscillations can be observed with this scheme.
We focus on modifying the usual $\Tone$ decay path from the (1,1) to (2,0) state.
In particular we focus on the (1,1)$\rightarrow$(2,1)$\rightarrow$(2,0) process (the (1,1)$\rightarrow$(1,0)$\rightarrow$(2,0) side of the triangle
also shows amplification effects under appropriate conditions and is not discussed further here). A rapid relaxation rate via this path
requires an electron of the appropriate spin to enter dot L rapidly (tunneling time $T_{L}$, typically $\sim$0.3 $\mu$s) from the left lead via the process 
(1,1)$\rightarrow$(2,1) and then for the electron in dot C to exit to the right lead (with tunneling time $T_{R}$) via the process 
(2,1)$\rightarrow$(2,0) as illustrated for a measurement at point 2 in \subfig{2}{a}, where $T_{R}\sim10 ns$, in the schematic \subfig{2}{d}. Consider the consequences of
slowing the second process down (a condition which can be trivially accomplished by reducing  the coupling to the right lead during the measurement
step of the pulse sequence i.e. making $T_{R} > \Tm$)  while maintaining a reasonably strong coupling to the other lead. The
(1,1)$\rightarrow$(2,1)  step can be made rapid compared to  $\Tm$ and $\Tone$ while the (2,1)$\rightarrow$(2,0) step can be comparable to both
$\Tone$ and $\Tm$, for example at point 3 in \subfig{2}{b} where $T_{R}\sim450 \mu$s. The result of this easily achieved adjustment
is that the contrast between the two states switches from  that between (1,1) and (2,0)  to that between (2,1) and (2,0). Since adding an electron to
the system has a much bigger  effect on the QPC charge detector than just transferring an electron between dots, this creates both a larger measurement
contrast and a larger signal to noise (these adjustments do not affect the level of intrinsic noise in the system).  Quantitatively, we can
estimate the effect by measuring the various QPC current step magnitudes between charge transitions. We  define an enhancement factor, E, by the ratio in the QPC
signal between (a)  transferring an electron from dot C to dot L i.e. (1,1)$\rightarrow$(2,0) and (b) removing one from dot C to the lead i.e.
(2,1)$\rightarrow$(2,0). In our device E is estimated to be $\simeq$-4  based on the QPC step heights. The minus sign indicates a predicted reversal of phase
in the LZS oscillations.  \subfig{2}{b} confirms this behaviour with LZS oscillations obtained using the above procedure. The device is set up  to have
the required barrier conditions for coherent oscillation enhancement. The measurement time in the (2,0) regime is 20 $\mu$s. LZS fringes can be 
observed in the stability diagram in the (2,0) regime. Since the period of rotation  depends on the energy separation of the qubit states and this in
turn depends on detuning, oscillations are expected  and observed as a function of detuning.  \subfig{3}{b} plots the derivative of the charge 
detector current while \subfig{3}{d} shows the raw charge detector current with a smooth plane removed and \subfig{3}{f} plots a line section from
the raw data.  The equivalent figures under unamplified conditions are given in \subfig{3}{a,c,e}. It is clear that oscillations with a much larger
amplitude are visible where (2,1) is employed as an intermediate state as described above (the amplifier region). 
The  oscillation amplitude in this region (see \subfig{3}{f}) is enhanced by a factor of about 4 compared to the unenhanced measurement
shown in \subfig{3}{e}.  This is consistent with the QPC step height estimate of E above. The oscillations in the enhanced region are
clearly phase shifted by $\pi$ relative to those in the unenhanced region (i.e white goes to black in \subfig{2}{b}) as expected based on the charge
detection argument above.  On varying $\Tm$ oscillations in  both regimes disappear at roughly the same value of $\Tm$ confirming that an anomalous
$\Tone$ value cannot be evoked to explain this amplification effect.

	In conclusion we have shown that one can use state sensitive relaxation to generate a larger contrast between 
qubit states at a charge detector. In our case we have demonstrated a signal to noise enhancement factor of 4 in
measurements of  S-$\Tplus$  qubit states. 

We acknowledge discussions with Bill Coish, Michel Pioro-Ladri\`ere, Ghislain Granger, Aash Clerk, and Guy Austing  
and O. Kodra for programming. A.S.S. acknowledges funding from NSERC and CIFAR.

\end{document}